\documentclass[aps,pra,twocolumn,superscriptaddress,showpacs]{revtex4-2}

\usepackage[utf8]{inputenc}

\usepackage{array}
\usepackage{amsmath}
\usepackage{graphicx}
\usepackage{epstopdf}
\usepackage{color}
\usepackage{amsfonts}
\usepackage{subfigure}
\usepackage{amssymb}
\usepackage{graphicx}
\usepackage{makecell}
\usepackage[colorlinks,citecolor=blue]{hyperref}
\usepackage{changes}
\def\ket#1{\mathinner{|{#1}\rangle}}

\newcommand{\onecolm}{
  \end{multicols}
  \vspace{-3.5ex}
  \noindent\rule{0.5\textwidth}{0.1ex}\rule{0.1ex}{2ex}\hfill
}
\newcommand{\twocolm}{
  \hfill\raisebox{-1.9ex}{\rule{0.1ex}{2ex}}\rule{0.5\textwidth}{0.1ex}
  \vspace{-4ex}
  \begin{multicols}{2}
}

\begin{document}

\title{The global phase diagram of the cluster-XY spin chain with dissipation}
\date{\today}

\author{Wei-Lin Li}
\affiliation {Key Laboratory of Atomic and Subatomic Structure and Quantum Control (Ministry of Education), Guangdong Basic Research Center of Excellence for Structure and Fundamental Interactions of Matter, School of Physics, South China Normal University, Guangzhou 510006, China}
\affiliation {Guangdong Provincial Key Laboratory of Quantum Engineering and Quantum Materials, Guangdong-Hong Kong Joint Laboratory of Quantum Matter, Frontier Research Institute for Physics, South China Normal University, Guangzhou 510006, China}

\author{Ying-Ao Chen}
\affiliation {DongGuan Experimental Middle School, Dongguan, 523007, China}

\author{Zheng-Xin Guo}
\email{zx\_guo@sjtu.edu.cn}
\affiliation{Wilczek Quantum Center and Key Laboratory of Artificial Structures and Quantum Control, School of Physics and Astronomy, Shanghai Jiao Tong University, Shanghai 200240, China}

\author{Xue-Jia Yu}
\email{xuejiayu815@gmail.com}
\affiliation{Department of Physics, Fuzhou University, Fuzhou 350116, Fujian, China}
\affiliation{Fujian Key Laboratory of Quantum Information and Quantum Optics,
College of Physics and Information Engineering,
Fuzhou University, Fuzhou, Fujian 350108, China}

\author{Zhi Li}
\email{lizphys@m.scnu.edu.cn}
\affiliation {Key Laboratory of Atomic and Subatomic Structure and Quantum Control (Ministry of Education), Guangdong Basic Research Center of Excellence for Structure and Fundamental Interactions of Matter, School of Physics, South China Normal University, Guangzhou 510006, China}
\affiliation {Guangdong Provincial Key Laboratory of Quantum Engineering and Quantum Materials, Guangdong-Hong Kong Joint Laboratory of Quantum Matter, Frontier Research Institute for Physics, South China Normal University, Guangzhou 510006, China}

\begin{abstract}
We study the ground-state phase diagram of a non-Hermitian cluster-XY spin chain in the language of free fermions. By calculating the second derivative of ground-state energy density and various types of order parameters, we establish the global ground-state phase diagram of the model, exhibiting rich quantum phases and corresponding phase transitions. Specially, the results reveal that the non-Hermitian cluster-XY model contains five different phases and two critical regions, i.e., ferromagnetic (FM), antiferromagnetic (AFM), symmetry-protected topological (SPT), paramagnetic (PM), Luttinger liquid-like phase, as well as critical region I and II. The order parameters and critical behaviors are investigated and the correctness of the theory is confirmed.
\end{abstract}

\maketitle


\section{INTRODUCTION}
With the ever-evolving ultracold atomic technology, optical lattice-based quantum simulation of ultracold atomic system has made rapid progress~\cite{BY2020,Meng2023,DJ1998,ZDW2016,ZSL2013,ZSL2007,ZSL2006,SLB2008}. Due to high controllability and purity, ultracold atomic systems are widely used to simulate phase transitions in condensed matter systems~\cite{Ji2014,Wang2023,Greiner2002,Zhang2021,SS2023,Xiao2021,CG2017,DJ1988,xie2020}. Recently, due to its unique SPT phase, Cluster spin model based on ultracold atoms in triangular lattices has received extensive attention~\cite{S2009,V2005,P2005,P2004,Becker2010}. A striking feature of this system is the coexistence of three-spin and two-spin couplings, and the competition between which will give rise to an exotic continuous quantum phase transition (QPT). In concrete terms, a phase transition from the SPT cluster phase to the symmetry-breaking phase occurs in the system~\cite{S2011,Ding2011,Son2011,NW2017,SMG2015,Guo2022,yu2402,yu2403,yu2404,yu2406,yu2022}. In the past few years, a series of models containing such continuous phase transitions have been studied, such as cluster-Ising model, cluster-XY model, etc. The relevant ground state phase diagrams have also been obtained one by one~\cite{Ding2011,Son2011,NW2017,SMG2015,Guo2022,yu2403}. Furthermore, some topological properties of the cluster spin model have been found, such as symmetry-protected edge modes at the gapped cluster SPT state and symmetry-enriched or topological nontrivial quantum critical points (QCPs)~\cite{yu2022,yu2402,yu2403,zhong2403,yu2404,yu2406}.

On the other hand, dissipation of the system is inevitable in almost all experimental platforms, be it a condensed matter platform or an artificial quantum simulation system~\cite{Li2019,HS2013,QL2022,SG2020,KY2021}, trapped ions~\cite{LLY2022,DCC2022,Zhang2022}. Therefore, dissipation is a factor that must be taken into account. It is also for this reason that many recent studies have discussed dissipative non-Hermitian systems. In addition to the experimental requirements, considering that dissipative non-Hermitian systems also have some unique properties that cannot be found in traditional Hermitian systems, for instance, non-Hermitian skin effect~\cite{QL2022,Longhi2024,Ma2024,Zhang2023}, non-Hermitian chiral properties~\cite{Xu2023,Shu2024,Yin2018}, exception points~\cite{Chen2017,Li2023,MAM2019}, spawning rings~\cite{Zhen2015}, mobility edge~\cite{lsz2404,LGJ2024}, etc. Recently, non-Hermitian physics has witnessed continuous progress and significant theoretical milestones~\cite{Yu2024prl,lsz2404,LSZyu2024,LHZ2023,Huang2022,Liu2021,Huang2021,Shen2019,He2020,He2021,He2022} including the hot topic of non-Hermitian topology~\cite{JRL2022,Liang2022,Gong2018,Liu2019,EM2019,EE2019,Luo2019,Zhang2020} and the nature of non-Hermitian exception points~\cite{Chen2017,Li2023,MAM2019}.

So far, although both SPT and non-Hermitian studies have come under the spotlight, few efforts are put to explore the properties of non-Hermitian SPT systems by combining the two. This work is devoted to the ground state properties and phase transitions of the dissipative cluster-XY model. We will construct a non-Hermitian cluster-XY model by introducing a complex field. Then, by the second derivative of ground state energy density, we show the non-Hermitian phase diagram.

The rest of this manuscript is organized as follows. In Sec.~\ref{sec:model}, we introduce the model and study the corresponding ground states by the analytical calculations. Then, we overview the phase diagram in Sec.~\ref{sec:phase}. We calculate the energy gap and various types of order parameters in Sec.~\ref{sec:order} to identify different phases. In Sec.~\ref{sec:variation}, we investigate the phase transitions and critical behaviors. We summarize this paper in Sec.~\ref{sec:summary}.


\section{MODEL AND analytical solution}
\label{sec:model}
\begin{figure}[tbh] \centering
\includegraphics[width=7.5cm]{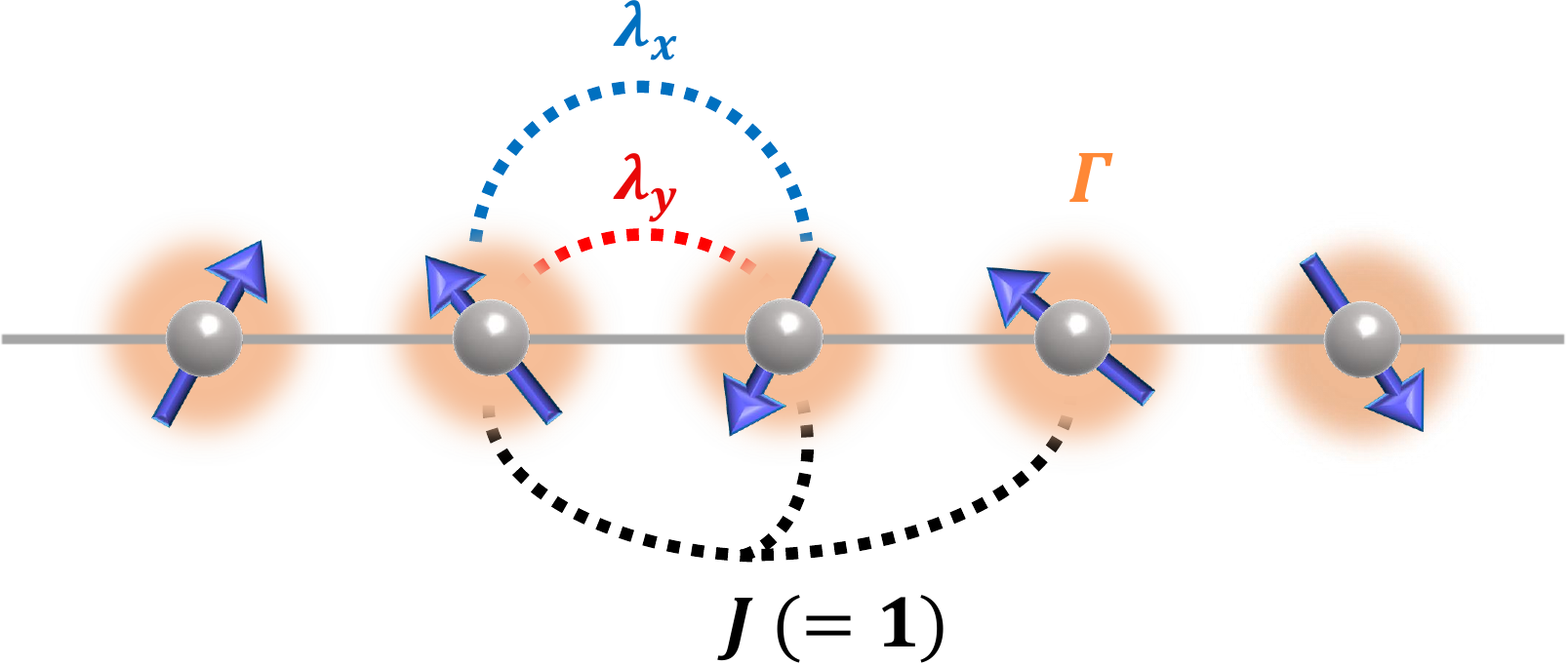}
\caption{Graphic demonstration of the cluster-XY model with dissipation, where $\lambda_{x}$ and $\lambda_{y}$ are the Ising exchange strength in $x$ and $y$ direction, respectively. $J$ is the strength of cluster term. $\Gamma$ is the dissipation strength. Throughout, we set $J=1$.}\label{1}
\end{figure}

We start with a dissipative cluster-XY model represented by an effective Hamiltonian with a complex field. The corresponding Hamiltonian reads
\begin{equation}\label{Hami}
\begin{aligned}
H=& -J\sum_{l=1}^N \sigma_{l-1}^x \sigma_l^z \sigma_{l+1}^x+\lambda_x \sum_{l=1}^N \sigma_l^x \sigma_{l+1}^x \\
& +\lambda_y \sum_{l=1}^N \sigma_l^y \sigma_{l+1}^y-\frac{i \Gamma}{2} \sum_{l=1}^N \sigma_l^u,
\end{aligned}
\end{equation}
where $\sigma^{\alpha}_{l}(\alpha=x,y,z)$ is the Pauli matrix of the $l$th spin. $\sigma^{u}=\begin{bmatrix}
1 & 0  \\
0 & 0
\end{bmatrix}$ denotes to loss or gain effect, which can be conveniently realized in optical systems and optical lattice ultracold atomic systems~\cite{Luo2019,P2012,Z2015}. Without loss of generality, we take $J=1$ as the unit of energy in following calculation. Experimentally, there are three controllable parameters, namely, the spin exchange strengths $\lambda_x$, $\lambda_y$ and the dissipation strength $\Gamma$ (see Fig.~\ref{1}).

One can transform Eq.~\eqref{Hami} into fermionic representation by conducting a Jordan-Wigner transformation, which is defined as
\begin{equation}
    \sigma_{l}^{z}=1-2c_{l}^{\dagger}c_{l},
\end{equation}
\begin{equation}
    \sigma_{l}^{+}=\prod_{j<l}(1-2c_{j}^{\dagger}c_{j})c_{l},
\end{equation}
where $c_l^{\dagger}$($c_l$) is the creation (annihilation) operator at site $l$. Then, one can perform Fourier transform
\begin{equation}
c_l=\frac{e^{-i \pi/4}}{\sqrt{N}} \sum_{k} e^{-i k l} c_k.
\end{equation}
Then, we obtain
\begin{equation}
\begin{aligned}
 H=&2 \sum_{k}\left[y_{k}\left(c_{k}^{\dagger} c_{-k}^{\dagger}+c_{-k} c_{k}\right)+z_{k}\left(c_{k}^{\dagger} c_{k}+c_{-k}^{\dagger} c_{-k}-1\right)\right],
\end{aligned}
\end{equation}
where $y_k=-\sin(2k)-(\lambda_{y}-\lambda_{x}) \sin(k)$ and $z_k=-\cos(2k)+(\lambda_{x}+\lambda_{y}) \cos(k)-\frac{i\Gamma}{4}$. By using Bogoliubov transformation,
\begin{equation}
    \gamma_{k}=u_{k}c_{k}+v_{k}c_{-k}^{\dagger},~\bar{\gamma}_{k}=u_{k}c_{k}^{\dagger}+v_{k}c_{-k}.
\end{equation}

Eventually, we get the diagonalized Hamiltonian
\begin{equation}
    H=\sum_{k} \Lambda_{k}(\bar{\gamma}_{k}\gamma_{k}-\frac{1}{2}),
\end{equation}
where
\begin{equation}\label{lambdak}
    \Lambda_k=\sqrt{y_{k}^2+z_{k}^2}.
\end{equation}
In this work, we define the ground state as the state with the minimum real part of $\Lambda_{k}$. The ground state of Eq.~\eqref{Hami} is
\begin{equation}
    \ket{G}=\frac{1}{\sqrt{N}}\prod_{k>0}[u_{k}-v_{k}c^{\dagger}_{k} c^{\dagger}_{-k}]\ket{0},
\end{equation}
where $N=\prod_{k>0}(|u_{k}|^2+|v_{k}|^2)$ is the normalization constant, $u_{k}=\frac{z_{k}-\Lambda_{k}}{C}$, $v_{k}=\frac{y_{k}}{C}$ and $C$ is a constant to satisfy $u_{k}^{2}+v_{k}^{2}=1$.

\section{PHASE DIAGRAM}\label{sec:phase}
\begin{figure}[tbhp] \centering
\includegraphics[width=7cm]{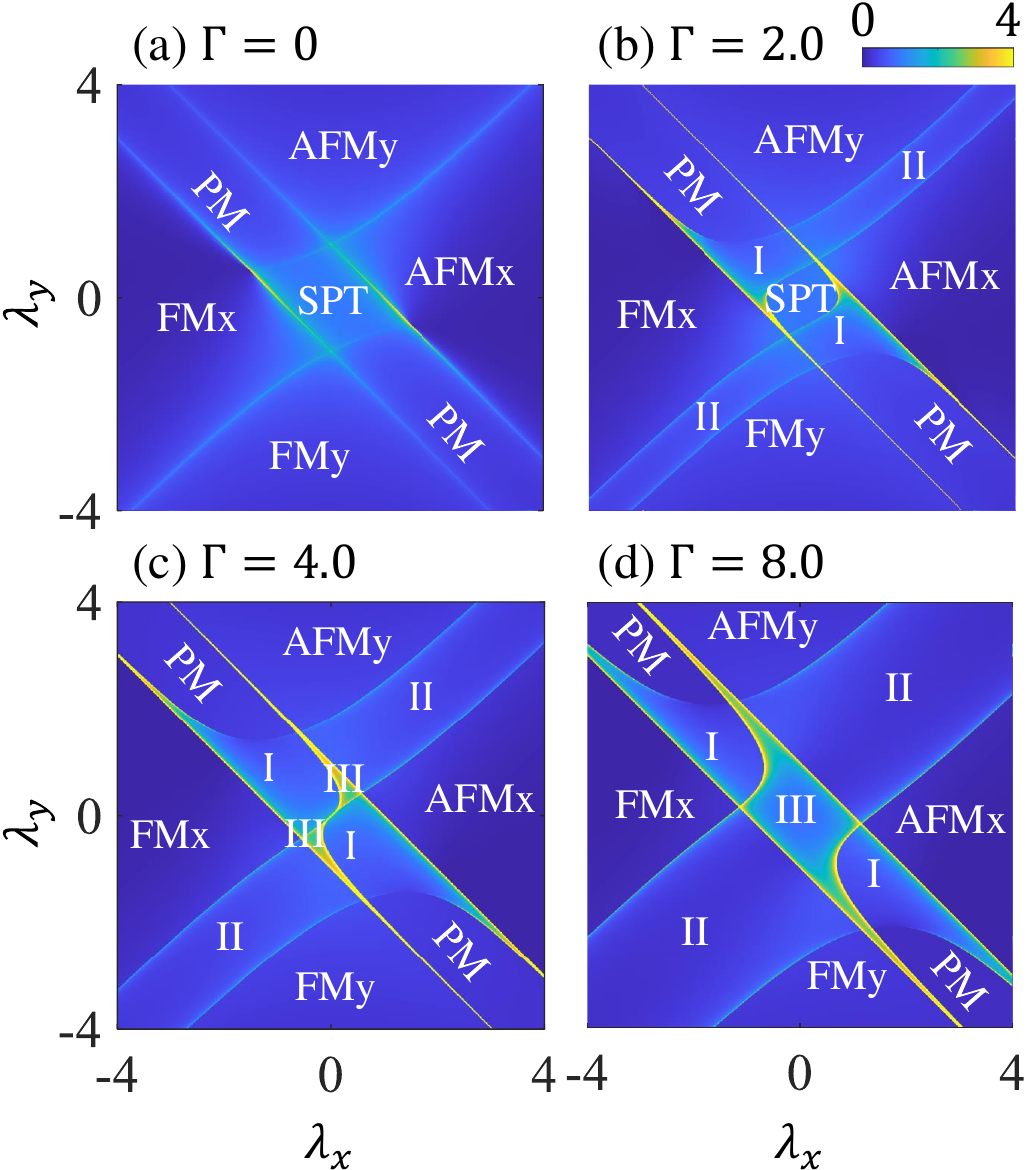}
\caption{The phase diagrams characterized by the real part of the second derivative of ground state energy density $-\frac{\partial^2 e_0}{\partial \lambda_{x}^{2}}$ for $\Gamma=0$ (a), $\Gamma=2.0$ (b), $\Gamma=4.0$ (c) and $\Gamma=8.0$ (d).}\label{2}
\end{figure}

The schematic phase diagram is provided in Fig.~\ref{2}. 
\begin{figure*}[tph] \centering
\includegraphics[width=17cm]{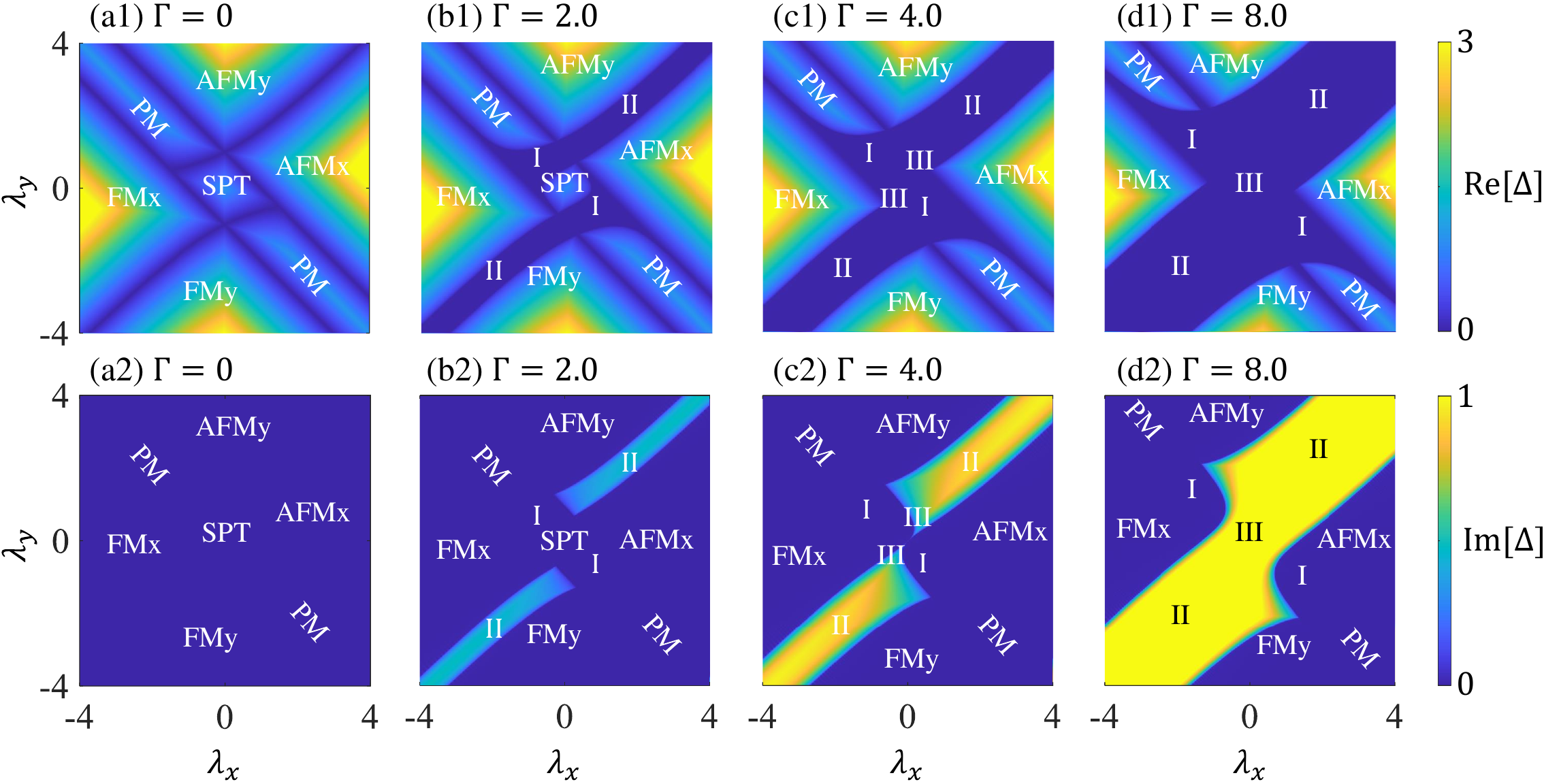}
\caption{The phase diagrams characterized by the real (top row) and imaginary (bottom row) parts of the energy gap for $\Gamma=0$, $2.0$, $4.0$, $8.0$.}\label{3}
\end{figure*}

Let's briefly outline the corresponding phase diagram and summarize the main findings. 

Under the condition of $\Gamma=0$, the model is reduced to the non-dissipative case, i.e., the standard Cluster-XY model, whose phase diagram is shown in Fig.~\ref{2}(a). On the other hand, the introduction of dissipation will bring about great changes in the phase diagram of the system. Specifically, when dissipation strength $\Gamma$ is weak, the SPT region will gradually shrink and two types of critical regions will appear in the system, namely, region I and region II~[see Fig.~\ref{2}(b)]. With a further increase in dissipation strength $\Gamma$, the SPT region will completely disappear, and a new type of phase III, the Luttinger liguid-like phase, will emerge from the system. After that, the region corresponding to phase III will increase with the ever-growing $\Gamma$ [see Figs.~\ref{2}(c)(d)].

The details of these emergent phases are briefly outlined below.

(1) In critical region I, both the real part and imaginary part of the energy gap are zero. As the distance $r$ increases, the string order parameter $|\mathcal{O}^{x}(r)|$ shows a power-law decay, the spin correlation function $|R_y|$ decays exponentially and $|R_x|$ oscillating decay as $r^{-a}$. 

(2) In critical region II, the energy gap is a pure imaginary number. With the distance $r$ increasing, the string order parameter $|\mathcal{O}^{x}(r)|$ decays exponentially, the spin correlation function $|R_x|,~|R_y|$ feature power-law decay. 

(3) In Luttinger liquid-like phase, the energy gap is a pure imaginary number. As the distance $r$ increases, the string order parameter $|\mathcal{O}^{x}(r)|$ shows a power-law decay and the spin correlation function $|R_y|$, $|R_x|$ present an oscillating decline as $r^{-a}$.

The corresponding properties of different phases are summarized in the following Tab.~\ref{tab1}. The phase transition both from critical region I to $\rm AFM_{x}$ and from Luttinger liquid-like phase to critical region II are first-order phase transitions. Furthermore, $\min|\Lambda_{k}|$ is an effective tool for detecting continuous phase transition in non-Hermitian cluster-XY model. In the following section, we will prove each of the above conclusions.

\section{Emergent gapless phases with dissipation}\label{sec:order}
Now, we explore the possible phases that appear in the phase diagram. Under the condition of $\Gamma=0$, the model is a standard cluster-XY model. By adjusting the parameters $\lambda_x,~\lambda_y$, the model contains four different phases, i.e., ferromagnetic (FM), antiferromagnetic (AFM), symmetry-protected topological (SPT), paramagnetic (PM)~\cite{MS2012}. However, when $\Gamma\neq0$, new phases emerge~[see Fig.~\ref{2}]. 
\begin{figure*}[tbhp] \centering
\includegraphics[width=16cm]{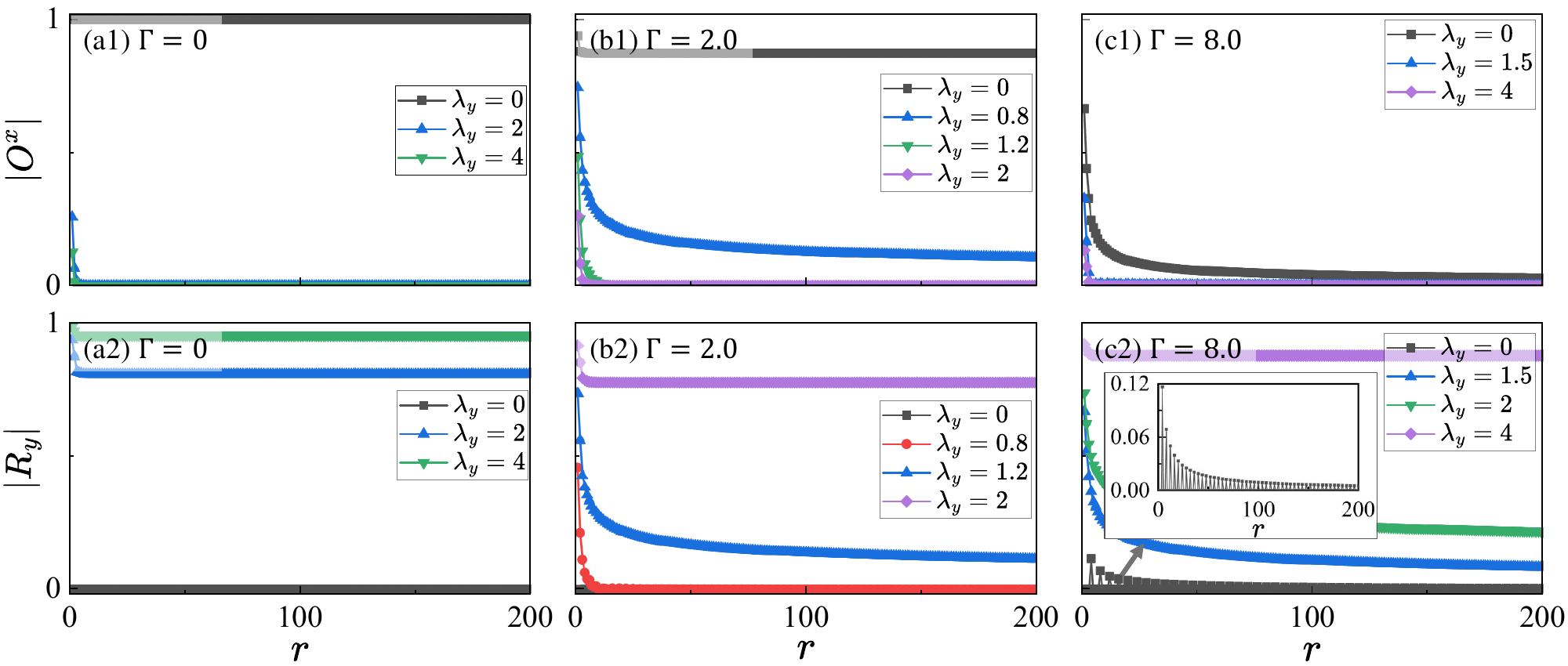}
\caption{(Color online). The long-distance behaviors of string order parameter $|\mathcal{O}^{x}|$ and spin correlation function $|R_{y}|$ for $\Gamma=0$ (a1)(a2), $\Gamma=2.0$ (b1)(b2), as well as $\Gamma=8.0$ (c1)(c2). The inset in (c2) shows that spin correlation function $|R_{y}|$ presents an oscillating decline as $r^{-0.7933}$. Throughout, $\lambda_{x}=0$.}\label{4}
\end{figure*}

In Hermitian cases, the minimum value of $\Lambda_{k}$ is defined as the energy gap, which is usually labeled as $\Delta$. The place where the energy gap closes ($\Delta=0$) is usually the critical point of phase transition. In non-Hermitian cases, however, since the value of $\Lambda_{k}$ is complex, the gap of the corresponding emergent phases is complex. Therefore, for the non-Hermitian case, we must examine both the real and imaginary parts of the energy gap. We plot the real (top row) and imaginary (bottom row) parts of the energy gap in Fig.~\ref{3}. Under the condition of $\Gamma=0$, the Re[$\Delta$] of different phase are all non-zero. The region I, II, III emerge and expand with an increasing $\Gamma$, and Re[$\Delta$] of these three emergent phases are zero [see Figs.~\ref{3}(b1)-(d1)]. The imaginary part, Im[$\Delta$] is zero in region I, whereas non-zero in region II and III. This is to say, the region I is a gapless phase, whereas the both region II and III are the imaginary-gapped phases~[see Figs.~\ref{3}(b2)-(d2)].

To identify each phase, we calculate the spin correlation function and string order parameter, which are two key quantities to study cluster spin model~\cite{S2011}. The spin correlation function is defined as
\begin{equation}
R_{\alpha}(r)=\left\langle\sigma_{j}^{\alpha} \sigma_{l}^{\alpha}\right\rangle,
\end{equation}
where $r=j-l$, $\alpha=x,~y,~z$. Then, one can obtain
\begin{equation}
\begin{aligned}
R_{x}(r) &=\left\langle\left(c_{j}-c_{j}^{\dagger}\right) \prod_{j<m<l}\left(1-2 c^{\dagger}_{m} c_{m}\right)\left(c_{l}^{\dagger}+c_{l}\right)\right\rangle \\
&=\left\langle B_{j} A_{j+1} B_{j+1} \ldots A_{l-1} B_{l-1} A_{l}\right\rangle,
\end{aligned}
\end{equation}
\begin{equation}
\begin{aligned}
R_{y}(r) &=(-1)^{r}\left\langle A_{j} B_{j+1} A_{j+1} \ldots B_{l-1} A_{l-1} B_{l}\right\rangle,
\end{aligned}
\end{equation}
where $A_j=c_j^{\dagger}+c_j$ , $B_j=c_j-c_j^{\dagger}$. There are pair contractions for $A_{j}$ and $B_{j}$, i.e.,
\begin{equation}
    \langle A_j A_l \rangle=\delta_{jl}+\frac{1}{\pi}\int_{0}^{\pi} dk (\frac{u_{k}v_{k}^{*}-u_{k}^{*}v_{k}}{|u_{k}|^{2}+|v_{k}|^{2}})\sin(kr),
\end{equation}
\begin{equation}
    \langle B_j B_l \rangle=-\delta_{jl}+\frac{1}{\pi}\int_{0}^{\pi} dk (\frac{u_{k}v_{k}^{*}-u_{k}^{*}v_{k}}{|u_{k}|^{2}+|v_{k}|^{2}})\sin(kr),
\end{equation}
\begin{equation}
\begin{aligned}
    \langle B_j A_l \rangle=-\frac{1}{\pi}\int_{0}^{\pi} dk (\frac{|u_{k}|^{2}-|v_{k}|^{2}}{|u_{k}|^{2}+|v_{k}|^{2}})\cos(kr) \\
    +\frac{1}{\pi}\int_{0}^{\pi} dk (\frac{u_{k}v_{k}^{*}-u_{k}^{*}v_{k}}{|u_{k}|^{2}+|v_{k}|^{2}})\sin(kr),
\end{aligned}
\end{equation}
where $r=l-j$~\cite{Lee2014}. Since both $R_{x}(r)$ and $R_{y}(r)$ contain a lot of operators, it is useful to write them in terms of the Pfaffian of a skew-symmetric matrix~\cite{wick1950}.

The string order parameter can be calculated in the same way as the spin correlation function. Then, we have
\begin{equation}
\mathcal{O}^{x}=\lim _{r \rightarrow \infty}(-1)^{r}\left\langle\sigma_{1}^{x} \sigma_{2}^{y}\left(\prod_{k=3}^{r-2} \sigma_{k}^{z}\right) \sigma_{r-1}^{y} \sigma_{r}^{x}\right\rangle.
\end{equation}

Similarly, by using $A_j=c_j^{\dagger}+c_j$ and $B_j=c_j-c_j^{\dagger}$, one can obtain
\begin{equation}
\mathcal{O}^{x}=\lim _{r \rightarrow \infty}\left\langle B_{1} B_{2} A_{3} B_{3} A_{4} B_{4}\ldots A_{r} B_{r} A_{r+1} A_{r+2}\right\rangle.
\end{equation}
Similarly, $\mathcal{O}^{x}$ can be converted to a Pfaffian of a skew-symmetric matrix. It is worth noting that $r$ should be as large as possible in the numerical calculation in order to approach the thermodynamic limit.

The string order parameter $\mathcal{O}^{x}$ tends to be a constant in the non-trivial cluster SPT phase, and decays exponentially in FM, AFM and PM phases. The spin correlation functions $|R_{x}(r)|$ and $|R_{y}(r)|$ tend to be a fixed non-zero constant in AFM phase along $x$ or $y$ direction, while they decay exponentially to zero in the disordered PM phase. 

Now, we exhibit a detailed analysis of the long-distance behaviors of order parameters.
\begin{figure*}[tbhp] \centering
\includegraphics[width=16cm]{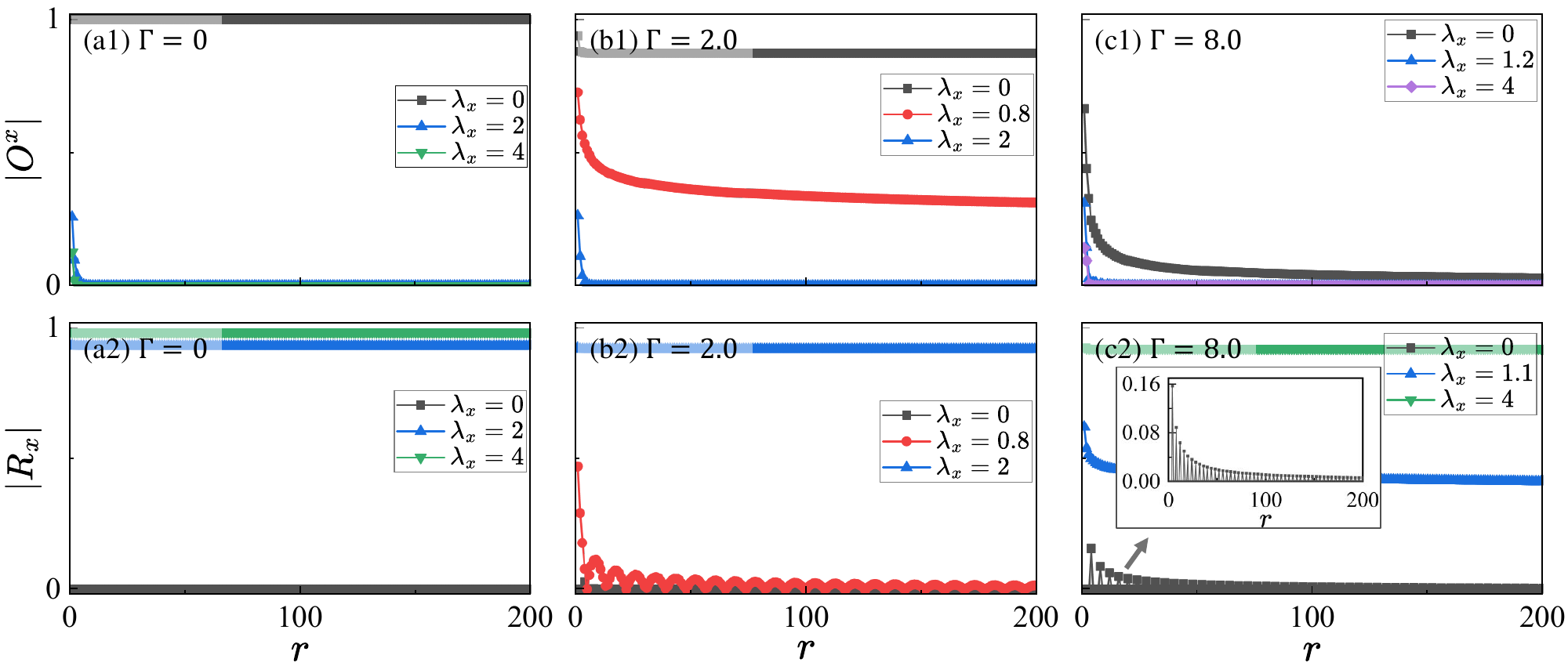}
\caption{(Color online). The long-distance behaviors of string order parameter $|\mathcal{O}^{x}|$ and spin correlation function $|R_{x}|$ for $\Gamma=0$ (a1)(a2) , $\Gamma=2.0$ (b1)(b2), and $\Gamma=8.0$ (c1)(c2). The inset in (c2) shows that spin correlation function $|R_{x}|$ presents an oscillating decline satisfying $r^{-0.8031}$. Throughout, $\lambda_{y}=0$.}\label{5}
\end{figure*}

Firstly, we set $\lambda_{x}=0$. Under the condition of $\Gamma=0$, one can find that when $\lambda_{y}>1$, the string order parameter $|\mathcal{O}^{x}|$ tends to be zero and the spin correlation function $|R_{y}|$ tends to be a constant, which means the corresponding region is $\rm AFM_{y}$ phase~[see Figs.~\ref{4}(a1) and (a2)]. Under the condition of $\lambda_{y}=0$, the string order parameter $|\mathcal{O}^{x}|$ tends to a constant and $|R_{y}|$ tends to be zero, which means corresponding region is the cluster SPT phase under such circumstance.

Under the condition of $\Gamma=2.0$, the order parameters' long-range behaviors become very different. For the case of $\lambda_{y}=2$, the string order parameter $|\mathcal{O}^{x}(r)|$ shows an exponential decay to be zero, whereas the spin correlation function $|R_{y}(r)|$ remains constant, indicating that the system resides in the $\rm AFM_y$ phase. In the middle region ($\lambda_y=0$), the string order parameter $|\mathcal{O}^{x}(r)|$ or the spin correlation function $|R_{y}(r)|$ becomes constant or tends to be zero in the long-distance limit, confirming that this region is the cluster SPT phase [see Figs.~\ref{4}(b1) and (b2)]. However, $|\mathcal{O}^{x}(r)|$ shows a power-law decay when $\lambda_y=0.8$ [see Fig.~\ref{4}(b1)], suggesting the presence of a quasi-long-range string order in region I. Then, as depicted in Fig.~\ref{4}(b2), one can observe that $|R_{y}(r)|$ features the power-law decay, implying the existence of quasi-long-range $\rm AFM_{y}$ order in region II. 

Under the condition of $\Gamma=8.0$, as shown in the Fig.~\ref{4}(c1), in region III ($\lambda_{y}=0$), the string order parameter $|\mathcal{O}^{x}(r)|$ shows a power-law decay as $r$ increases, suggesting the existence of a quasi-long-range string order. And one can observe that the spin correlation function $|R_y|$ presents an oscillating decline as $r^{-a}$ in region III [see Fig.~\ref{4}(c2)].

Secondly, we set $\lambda_{y}=0$. When $\Gamma=0$, as can be seen in Figs.~\ref{5}(a1) and (a2), when $\lambda_{x}>1$, the string order parameter $|\mathcal{O}^{x}|$ or the spin correlation function $|R_x|$ tends to be zero or remains a constant value, confirming that the region is $\rm AFM_x$ phase. As depicted in Fig.~\ref{5}(b1), the string order parameter $|\mathcal{O}^{x}(r)|$ also shows a power-law decay in region I ($\lambda_x=0.8$). Interestingly, in region I, $|R_{x}(r)|$ presents an oscillating decline as the distance $r$ increases [see Fig.~\ref{5}(b2)], which is consistent with its behavior at the SPT-PM phase transition point when $\Gamma=0$ [see Fig.~\ref{7}(a)]. So region I is a critical region which emerges from SPT-PM phase transition line with an increasing disspative strength $\Gamma$.

When $\Gamma=8$, according to Fig.~\ref{5}(c2), one can observe that $|R_{x}(r)|$ shows the power-law decay in the critical region II, which is consistent with its behavior at the $\rm AFM_x$-$\rm AFM_y$ phase transition point when $\Gamma=0$ [see Fig.~\ref{7}(b)]. So one can consider that region II is a critical region emerging from $\rm AFM_x$-$\rm AFM_y$ phase transition line with an increasing disspative strength $\Gamma$. As depicted in the inset of Fig.~\ref{5}(c2), the spin correlation function $|R_x|$ presents an oscillating decline as $r^{-a}$ in region III. Combining with the long-distance behavior of $|R_y|$ shown in Fig.~\ref{4}(c2), we define that the region III is a Luttinger liquid-like phase. In addition, we also investigate the long-distance behaviors of correlation functions in PM phase (see Appendix~\ref{sec:appA} for details).

We summarize the corresponding properties of the energy gap and correlation functions of different phases and critical regions in Tab.~\ref{tab1}.
\begin{ruledtabular}
\begin{table*}[hbtp]
\caption{The energy gap and long-distance behaviors of order parameters in different phases.}
\label{tab1}
\begin{tabular}{cccccc}
& $\Delta$ & $|\mathcal{O}^x|$ & $|R_x|$ & $|R_y|$  \\
\hline
 SPT & real & constant & 0 & 0 \\
 PM & real & exponential decay & exponential decay & exponential decay \\
 $\rm FM_{x}$($\rm AFM_{x}$) & real & exponential decay & constant & exponential decay \\
 $\rm FM_{y}$($\rm AFM_{y}$) & real & exponential decay & exponential decay & constant \\
 Critical region I & 0 & power-law decay & oscillating decay as $r^{-a}$ & exponential decay\\
 Critical region II & imaginary & exponential decay & power-law decay & power-law decay\\
 Luttinger liquid-like phase & imaginary & power-law decay & oscillating decay as $r^{-a}$ & oscillating decay as $r^{-a}$\\ 
\end{tabular}
\end{table*}
\end{ruledtabular}
\begin{figure}[tph] \centering
\includegraphics[width=8cm]{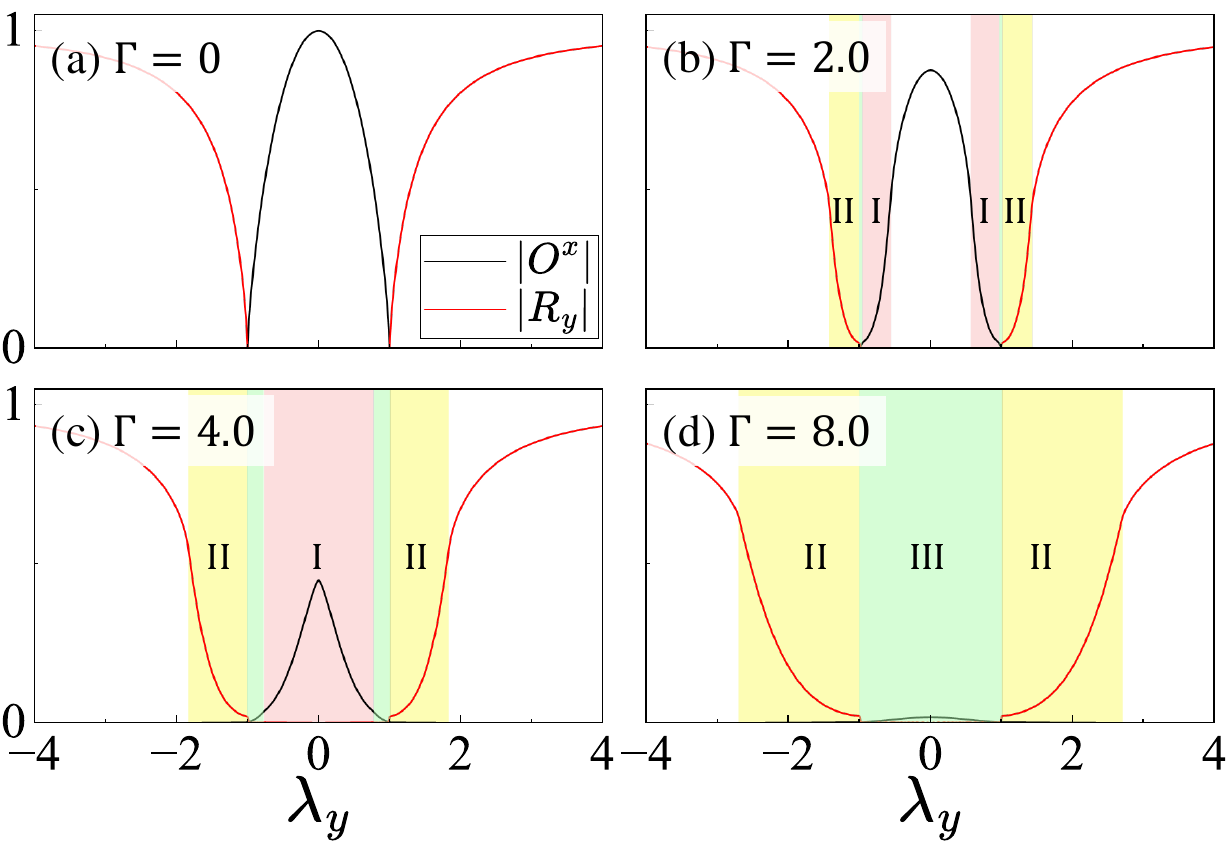}
\caption{(Color online). The numerical results of string order parameter $|\mathcal{O}^{x}|$, spin correlation function $|R_{y}|$ with respect to $\lambda_{y}$ for $\Gamma=0$ (a), $\Gamma=2.0$ (b), $\Gamma=4.0$ (c) and $\Gamma=8.0$ (d). The red, yellow, green shading correspond to critical region I, II, Luttinger liquid-like phase, respectively. Throughout, $\lambda_{x}=0$, $r=1000$.}\label{6}
\end{figure}

On the other hand, we set $\lambda_{x}=0$, $r=1000$, and study the distribution of the order parameters under different disspative strengths. In the Hermitian case [see Fig.~\ref{6}(a)], when $\lambda_{y}>1$ ($\lambda_{y}<1$), $|R_{y}|$ is non-zero, indicating that the system resides in the $\rm AFM_{y}$($\rm FM_{y}$) phase in such a parameter region. When $-1<\lambda_{y}< 1$, $|\mathcal{O}^{x}|$ is non-zero, suggesting that the region is the cluster SPT phase. Then we investigate the non-Hermitian case. As depicted in Fig.~\ref{6}(b), when $\Gamma=2.0$, one can observe that $|R_{y}|$ is a constant in critical region II and $|\mathcal{O}^{x}|$ is a constant in critical region I. As dissipative strength $\Gamma$ increases from $2.0$ to $4.0$, the SPT phase narrows, while critical region I, critical region II and Luttinger liquid-like phase expand [see Figs.~\ref{6}(b) and (c)]. When $\Gamma$ increases to $4.0$, the region of SPT phase disappears completely [see Fig.~\ref{6}(c)]. As the disspative strength $\Gamma$ further increases, the critical region II and Luttinger liquid-like phase continuously expand [see Fig.~\ref{6}(d)]. Additionally, we investigate the distribution of the order parameters under different disspative strengths when $\lambda_{y}=0$ (see Appendix~\ref{sec:appC} for details).

\section{Phase transitions and critical behaviors}
\label{sec:variation}

After delineating all the quantum phases in the phase diagram, we shift our focus to the more intriguing QPTs between these phases. In this section, we are ready to study phase transitions and critical behaviors. According to Eq.~\eqref{lambdak}, the ground state energy density can be defined as
\begin{equation}
e_0=-\frac{2}{N}\sum_{k}\Lambda_k=-\frac{1}{\pi}\int_{0}^{\pi}\sqrt{y_{k}^{2}+z_{k}^{2}}~dk,
\end{equation}
and we can easily obtain the second derivative of $e_0$ with respect to $\lambda_x$, i.e., $-\frac{\partial^2 e_0}{\partial \lambda_{x}^{2}}$. By means of the second derivative of $e_0$ calculations, one can observe that the second derivative of the ground state energy density $-\frac{\partial^2 e_0}{\partial \lambda_{x}^{2}}$ becomes sharper at critical points. When $\Gamma\neq0$, some new phase transition lines emerge. One can observe that critical region I and critical region II emerge from the transition line of SPT-PM and $\rm AFM_{y}-AFM_{x}$ phase in Hermitian case. So in order to explain the properties of critical region I and critical region II, we investigate the properties of correlation functions at the critical points of SPT-PM and $\rm AFM_{y}-\rm AFM_{x}$ transitions when $\Gamma=0$. 

\begin{figure}[tbhp] \centering
\includegraphics[width=8cm]{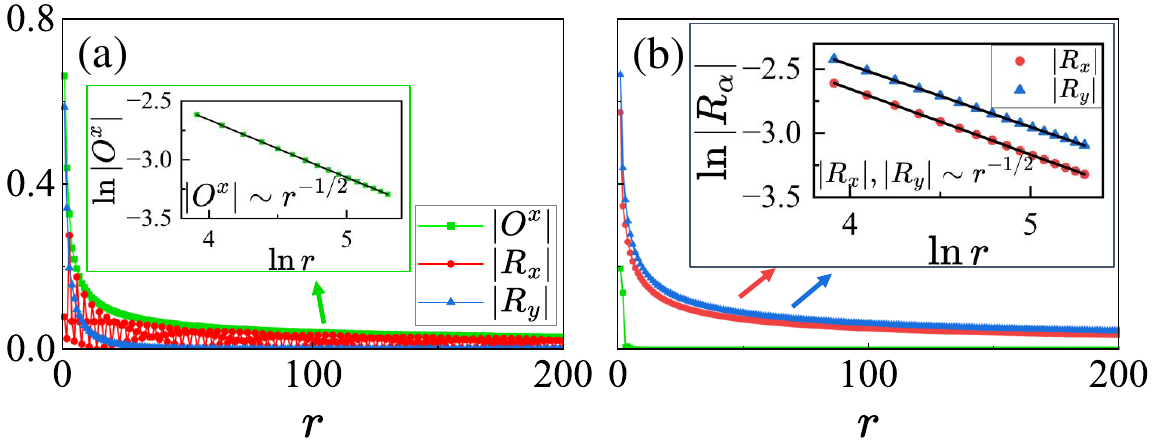}
\caption{(Color online). The long-distance behaviors of string order parameter $|\mathcal{O}^{x}|$ and spin correlation function $|R_{x}|$, $|R_{y}|$ for $\lambda_x=0.1,~\lambda_y=-0.9512$ (a), $\lambda_x=3.0,~\lambda_y=3.3030$ (b). Throughout, $\Gamma=0$.}\label{7}
\end{figure}

The numerical results are depicted in Fig.~\ref{7}. As shown in Fig.~\ref{7}(a), the string order parameter $|\mathcal{O}^x|$ shows a power-law decay, the spin correlation function $|R_x|$ shows oscillating decay as $r^{-a}$ and $|R_y|$ decays exponentially at the critical point of SPT-PM transitions. These long-distance behaviors of the correlation functions are the same as these in critical region I. The inset shows that the slope of curves in ln-ln plot is close to $-1/2$, implying that the critical exponent $\eta$ of SPT-PM transitions is $1/2$ [see Fig.~\ref{7}(a)]. Then, one can observe that the spin correlation function $|R_x|,~|R_y|$ show power-law decay and the string order parameter $|\mathcal{O}^{x}|$ exhibits an exponential decay at the critical point of $\rm AFM_{y}-\rm AFM_{x}$ transitions [see Fig.~\ref{7}(b)]. These long-distance behaviors of the correlation function are consistent with the properties of them in the critical region II. The inset shows that the slope of curves in ln-ln plot is close to $-1/2$, implying that the critical exponent $\eta$ of $\rm AFM_{y}-AFM_{x}$ transitions is $1/2$. 

\begin{figure}[tbhp] \centering
\includegraphics[width=8cm]{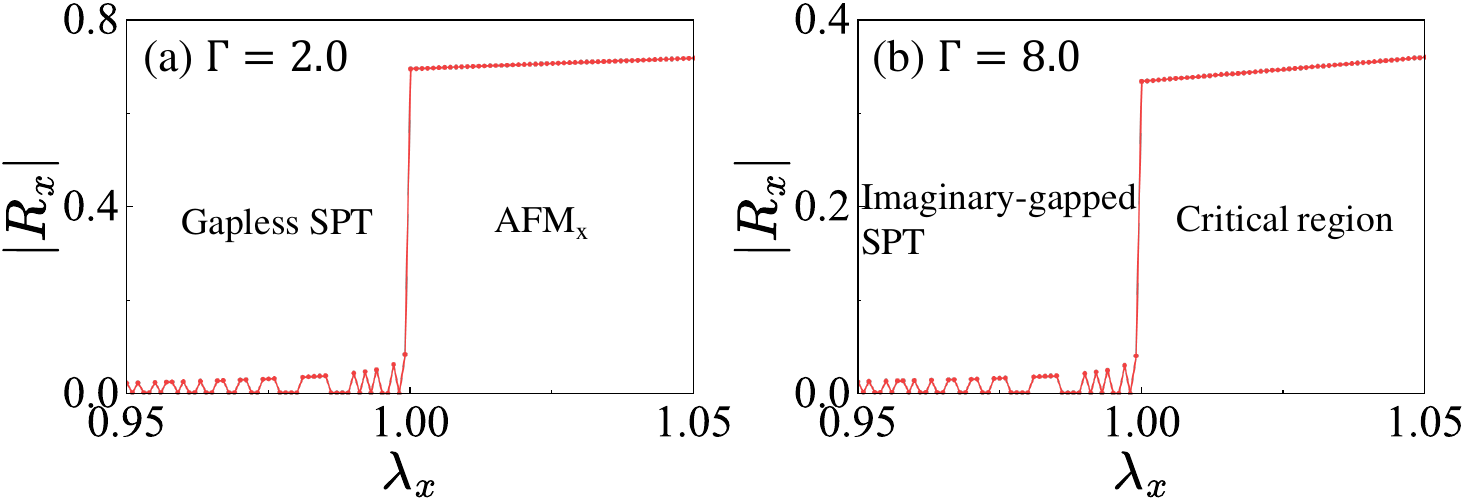}
\caption{(Color online). The spin correlation
function $|R_x|$ with respect to $\lambda_{x}$ for $\Gamma = 2.0$ (a) and $\Gamma = 8.0$ (b). Throughout, $\lambda_{y}=0$, $r = 1000$.}\label{8}
\end{figure}

\begin{figure}[tbhp] \centering
\includegraphics[width=8cm]{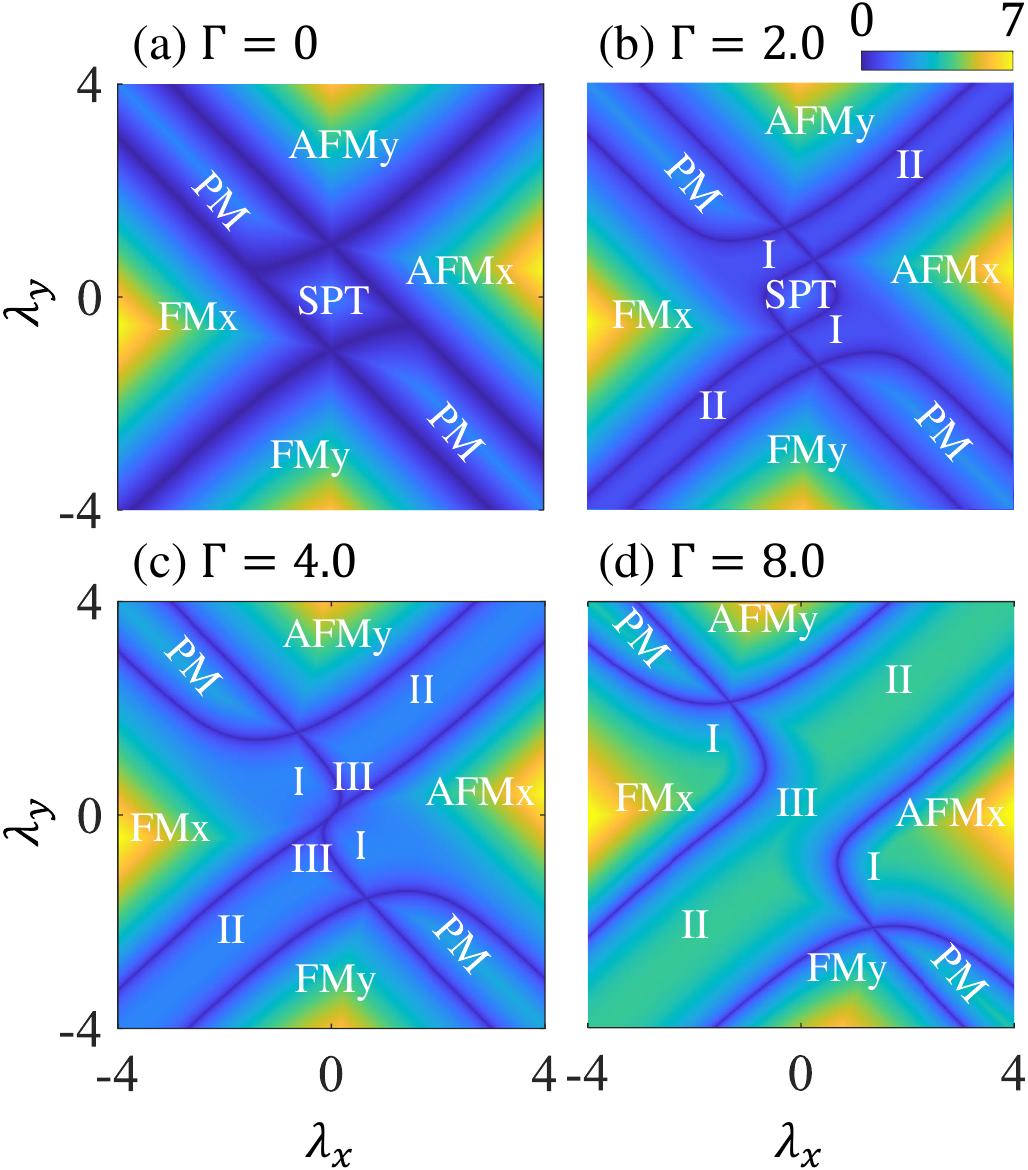}
\caption{The phase diagrams characterized by $\min |\Lambda_{k}|$ for $\Gamma=0$ (a), $\Gamma=2.0$ (b), $\Gamma=4.0$ (c) and $\Gamma=8.0$ (d). }\label{9}
\end{figure}

On the other hand, in order to determine whether the phase transition is first-order phase transition or not, we study the scaling behaviors of the order parameters at the QCPs. The numerical results are presented in Fig.~\ref{8}. One can observe that when $\Gamma=2.0$, the jump in the spin correlation function $|R_{x}|$ at critical region I-$\rm AFM_{x}$ transition and another jump at Luttinger liquid-like phase-critical region II transition, which indicate that the transitions of both critical region I-$\rm AFM_{x}$ and Luttinger liquid-like phase-critical region II are first-order phase transition [see Fig.~\ref{8}]. Combining with the previous numerical results of the energy gap [see Fig.~\ref{3}], one can discover that both the real energy gap and imaginary energy gap closing points are not corresponding to the continuous phase transition. Interestingly, one can discover that the points of $\min |\Lambda_{k}|=0$ correspond to the continuous phase transitions in our system [see Fig.~\ref{9}].
\section{Summary}
\label{sec:summary}
In summary, we investigate the effect of disspation on the phase diagram of the cluster-XY model. By means of the second derivative of ground state energy density calculation, one can observe that the introduction of established fields can destroy the SPT phase and emerge three novel phases. By calculating the energy gap and order parameters, we obtain the properties of different phases and the critical behaviors at the points of phase transitions. In critical region I, the string order parameter $|\mathcal{O}^{x}|$ exhibits power-law decay, the spin correlation function $|R_y|$ decays exponentially and $|R_x|$ presents an oscillating decay as $r^{-a}$, which are consistent with the critical behaviors of SPT-PM transition. Different from critical region I, the string order parameter $|\mathcal{O}^{x}(r)|$ decays exponentially, the spin correlation function $|R_x|,~|R_y|$ feature power-law decay in critical region II, which are consistent with the critical behaviors of $\rm AFM_{x}$-$\rm AFM_{y}$ transition. In Luttinger liquid-like phase, the string order parameter $|\mathcal{O}^{x}(r)|$ satisfies power-law decay and the spin correlation function $|R_y|$, $|R_x|$ presents an oscillating decay as $r^{-a}$. Along with the emergent phases, the transition of critical region I-$\rm AFM_{x}$ and Luttinger liquid-like phase-critical region belong to first-order phase transition. Different from Hermitian system, both real part and imaginary part of energy gap closing points are not corresponding to the phase transition points completely. Interestingly, the continuous phase transition occur with $\min|\Lambda_{k}|=0$. Our series of theoretical work (Ref.\cite{Guo2022} and this paper) will be a constant push to the ever-deepening research on novel phases and phase transitions in cluster spin system.

\acknowledgments
We sincerely thank D.-W. Zhang for helpful discussions. This work was supported by the National Key Research and Development Program of China (Grant No.~2022YFA1405300). X.-J. Yu was supported by the National Natural Science Foundation of China (Grant No.12405034).\\

\appendix

\section{Details about correlation functions in PM phase}
\label{sec:appA}
\begin{figure}[tbhp] \centering
\includegraphics[width=8cm]{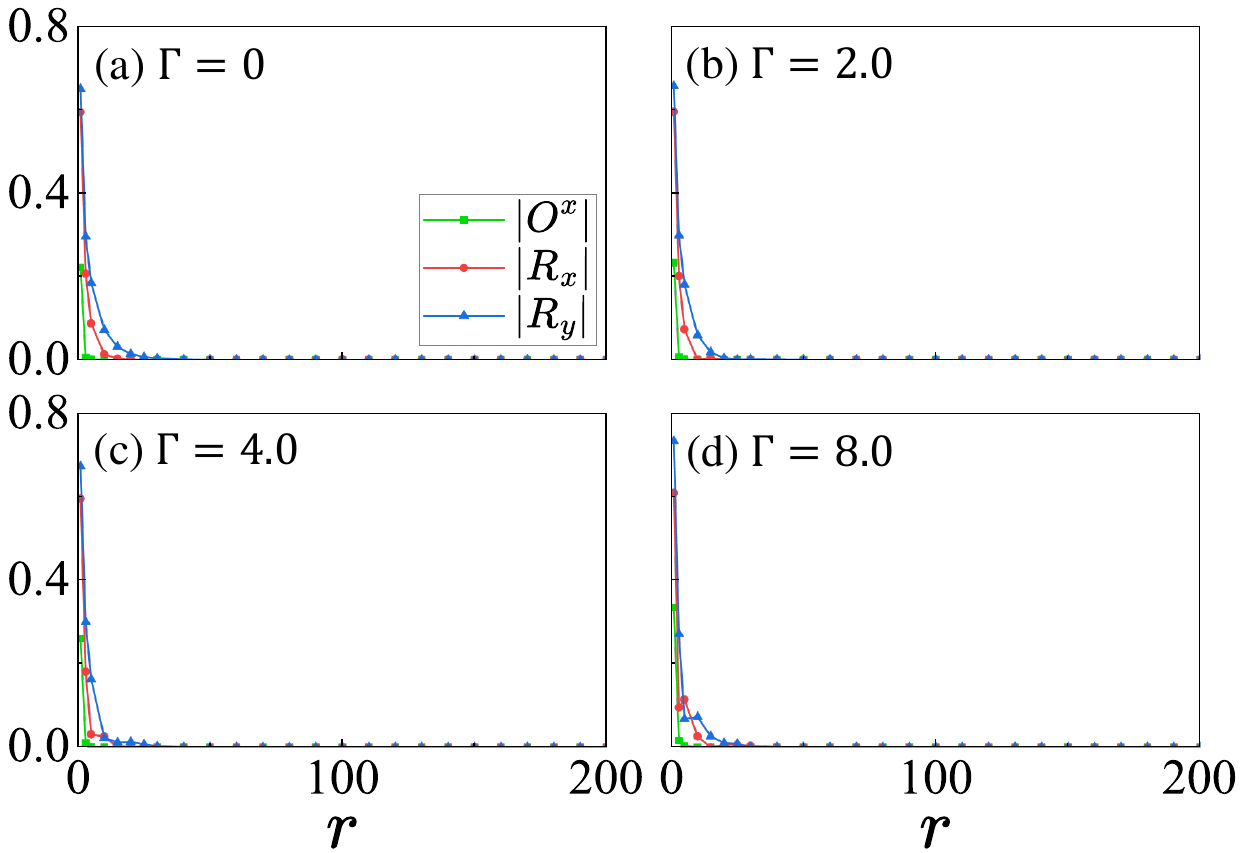}
\caption{(Color online). The long-distance behaviors of string order parameter $|\mathcal{O}^{x}|$ and spin correlation function $|R_{x}|$, $|R_{y}|$ for $\Gamma=0$ (a), $\Gamma=2.0$ (b), $\Gamma=4.0$ (c) and $\Gamma=8.0$ (d). Throughout, $\lambda_x=-3$, $\lambda_y=3$.}\label{10}
\end{figure}
In this section, we provide the properties of correlation functions in PM phase. As shown in Fig.~\ref{10}, the results reveal that the string order parameter $\mathcal{O}^x$ and the spin correlation function $|R_x|,~|R_y|$ decay exponentially with the distance $r$ increasing in PM phase.

\section{The distribution of order parameters versus $\lambda_{x}$}
\label{sec:appC}
\begin{figure}[tbhp] \centering
\includegraphics[width=8cm]{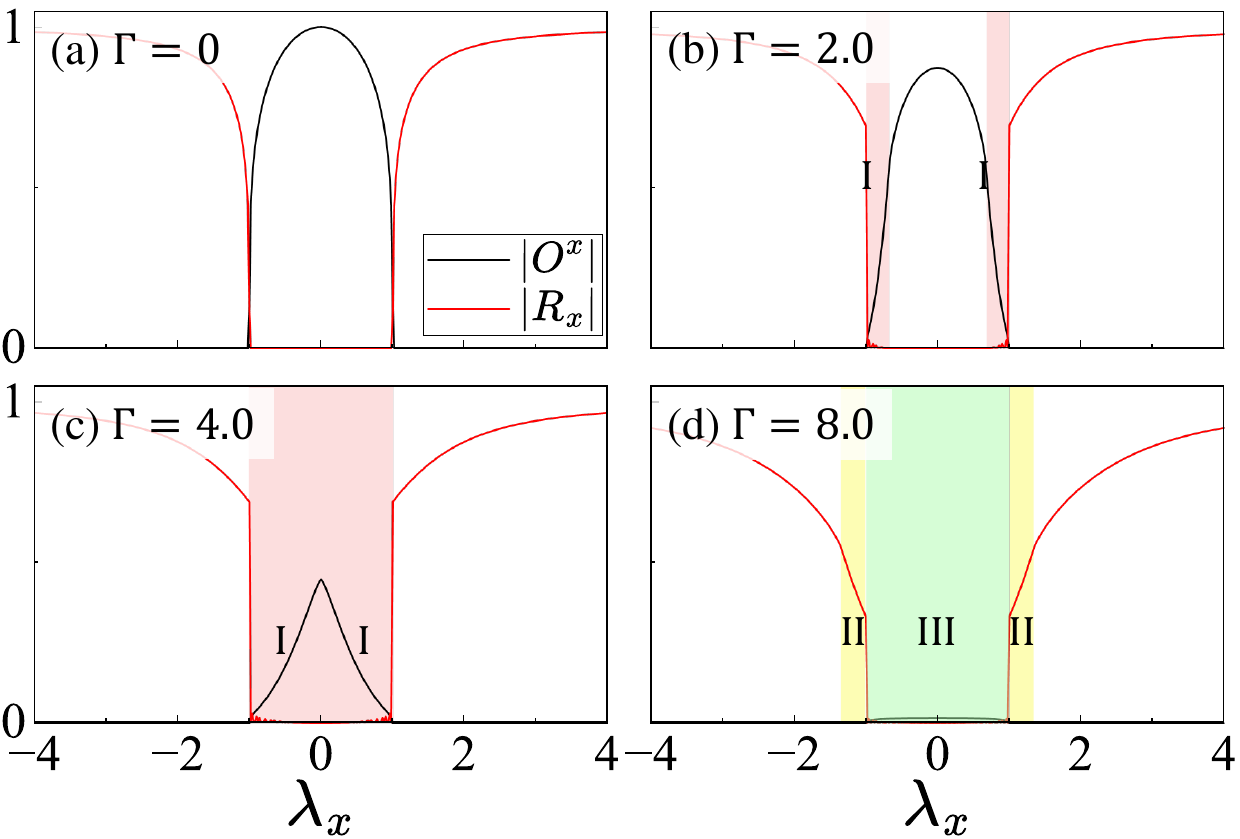}
\caption{(Color online). The numerical results of string order parameter $|\mathcal{O}^{x}|$, spin correlation function $|R_{x}|$ with respect to $\lambda_{x}$ for $\Gamma=0$ (a), $\Gamma=2.0$ (b), $\Gamma=4.0$ (c) and $\Gamma=8.0$ (d). The red, yellow, green shading correspond to critical region I, II, Luttinger liquid-like phase, respectively. Throughout, $\lambda_{y}=0$, $r=1000$.}\label{11}
\end{figure}

In this section, we present additional data on the distribution of the order parameters of $|\mathcal{O}^{x}|$, $|R_{x}|$ versus $\lambda_x$ when $\lambda_{y}=0$. In the Hermitian case ($\Gamma=0$), one can find that $|R_{x}|$ is non-zero when $\lambda_{x}>1$ ($\lambda_{x}<-1$), and $|\mathcal{O}^{x}|$ is non-zero when $-1<\lambda_{x}<1$ [see Fig.~\ref{11}(a)], which reveal that these regions are $\rm AFM_x(FM_x)$ and SPT phase. 

Then, we investigate the non-Hermitian case. Under the condition of $\Gamma=2$, as shown in Fig.~\ref{11}(b), in critical region I, $|\mathcal{O}^{x}|$ is non-zero. As the dissipation intensity increases, the range of each phase changes. When the dissipation intensity increases to $\Gamma=4$, the phase with SPT completely disappears, which is in agreement with the behavior in the phase diagram [see Fig.~\ref{11}(c)]. Under the condition of $\Gamma=8.0$, it can be seen that the spin correlation function $|R_{x}|$ is a limited value in the critical region II and the string order parameter $|\mathcal{O}^{x}|$ is non-zero in Luttinger liquid-like phase [see Fig.~\ref{11}(d)].

\bibliography{main}

\end{document}